# Towards pump-probe experiments of defect dynamics with short ion beam pulses

T. Schenkel[1], S. M. Lidia[1], C. D. Weis[1], W. L. Waldron[1], J. Schwartz[1], A. M. Minor[2,3], P. Hosemann[2,4], and J. W. Kwan[1]

[1]Accelerator and Fusion Research Division, Lawrence Berkeley National Laboratory, 1 Cyclotron Road, 5R121, Berkeley, CA 94720, USA

[2]Materials Science Division, Lawrence Berkeley National Laboratory, 1 Cyclotron Road, Berkeley, CA 94720, USA

[3]Department of Materials Science and Engineering, University of California, Berkeley, CA 94720, USA

[4]Nuclear Engineering Department, University of California, Berkeley, CA 94720, USA

A novel, induction type linear accelerator, the Neutralized Drift Compression eXperiment (NDCX-II), is currently being commissioned at Berkeley Lab. This accelerator is designed to deliver intense (up to $3 \times 10^{11}$ ions/pulse), 0.6 to ~600 ns duration pulses of 0.13 to 1.2 MeV lithium ions at a rate of about 2 pulses per minute onto 1 to 10 mm scale target areas. When focused to mm-diameter spots, the beam is predicted to volumetrically heat micrometer thick foils to temperatures of ~30,000 °K. At lower beam power densities, the short excitation pulse with tunable intensity and time profile enables pump-probe type studies of defect dynamics in a broad range of materials. We briefly describe the accelerator concept and design, present results from beam pulse shaping experiments and discuss examples of pump-probe type studies of defect dynamics following irradiation of materials with intense, short ion beam pulses from NDCX-II.

## 1.    Introduction

Intense, short pulses of energetic ions are highly desirable for studies of warm dense matter, high energy density physics experiments with volumetrically heated targets, and for studies of defect dynamics in solids [1, 2].  Irradiation with short ion beam pulses generates defects in a narrow time window and their diffusion and recombination dynamics can then be studied in time resolved "pump – probe" type experiments.  Here, the short ion beam pulse acts as the "pump" which can be followed by a suitable "probe" beam, such as an x-ray pulse, which can track a selected defect signature.  Combining a short pulse ion beam capability with pulsed probe beams from an x-ray free electron laser (FEL) was recently proposed for the study of radiation effects in nuclear materials by Froideval et al [1].  Suitable driver beams can be formed through longitudinal compression of space-charge dominated ion beams where short pulses have been achieved by imposing head-to-tail velocity tilts to drifting ion beams [2-4].  High drift compression factors are reached when ion beams traveled through a neutralizing plasma column during drift compression.  Earlier, a 25 mA, 300 keV $K^+$ beam was compressed 50 to 90 fold, yielding an intense ~3 ns long pulse with a beam spot size of ~1 mm$^2$ [4].  In order to achieve uniform heating of micrometer thick foil targets to temperatures of 2 – 3 eV, we are currently implementing this beam compression concept with increased ion beam energy (up to 1.2 MeV) for lithium ions [3, 5]. The induction linac based accelerator affords a high degree of flexibility in ion beam pulse shaping. In an intensity regime well below the onset of intense target heating and hydrodynamic motion, this flexibility enables the study of defect dynamics in solids and other materials (including soft matter and liquids).  Here, targets can be exposed to short ion beam pulses and resulting defect structures can be probed with time-resolved *in situ* or with *ex situ* methods. This article is structured as follows: We first present results from ion beam pulse control experiments at NDCX-II in Section 2.  We then outline concepts for pump-probe experiments for defect dynamics studies and present results from short pulse implantation of lithium ions into silicon (Section 3), followed by an outlook and conclusions.

## 2.        Ion beam pulse shaping experiments

        The NDCX-II accelerator is assembled from a modular cell structure, where pulsed induction cells are iterated with diagnostic cells and drift cells, all embedded with pulsed solenoid magnets [2-5].  The NDCX-II induction linac structure has a length of 12 m (Figure 1). Each active induction cell presents either a purely accelerating voltage or a time-varying, ramped voltage to the non-relativistic ion beam.  Additional inactive (i. e. unpowered) induction cells as well as diagnostic cells are necessary to allow the velocity-modulated ion beam to complete its initial compression phase.  Here, we report results from beam compression experiments with a 27-cell configuration that utilizes seven active induction cells, six diagnostics cells and 14 inactive ("drift") cells.  Lithium ions are extracted in ~600 ns long pulses from a large area (10.9 cm diameter), lithium doped, thermionic alumino-silicate emitter with an emitted current density of ~1 mA/cm$^2$.  The ion source operates at a surface temperature of ~1200°C and is radiatively heated by a filament with power consumption of ~4 kW.  Ions are extracted, accelerated and focused within the injector by a triode structure [6].  Emitted ion beam currents are ~50 mA.

The same source type has also been used to form intense pulses of other alkali ions, e. g. $K^+$ [4]. Lithium ions are generated and then injected into the first acceleration cell with an initial kinetic energy of up to 140 keV. In the first cell, we apply a voltage that accelerates the beam by another 20 keV. Using several more ramped-voltage induction cells, we have achieved compression of lithium ion pulses from ~600 ns to 20 ns (FWHM) with this 27-cell beamline configuration (Figure 2). Radial beam focusing is enabled by pulsed solenoid magnets within each cell with magnetic fields ranging from ~0.8 to 2.5 T. In the example in Figure 2, the peak intensity is 918 mA with about 13 nC of charge in the 20 ns pulse (FWHM), corresponding to $8\times10^{10}$ ions/pulse. Here, the un-neutralized beam pulses are not yet strongly focused laterally and beam spots have diameters of about 10 mm. Adaptation of a plasma chamber for space charge neutralization and an 8 T pulsed solenoid magnet will focus the beam to a spot with a diameter of ~1 mm [3-5], as well as continue longitudinal compression of the beam pulse to its ultimate pulse duration of ~0.6 ns. However, for studies of defect dynamics, it is desirable to avoid intense sample heating and the energy fluence can be controlled by varying the beam spot size or the number of ions present in a beam pulse. In Figure 3 we show an image of the beam induced intensity distribution of light emission from a 0.1 mm thick aluminum oxide scintillator. Ionoluminescence from the aluminum oxide target is imaged in transmission with a fast (~30 ns), gated CCD camera. In the example shown in Figure 3, the beam energy was 160 keV, the total charge per pulse was 28 nC and the beam spot had a diameter of 12 mm (FWHM). The resulting peak energy fluence was 1.8 mJ/cm$^2$. We summarize the beam parameters of NDCX-II in Table 1, comparing the current status to the design goals. The design goals of NDCX-II will enable volumetric heating of thin foils to a few eV temperatures. For studies of defect dynamics, the tuning flexibility of the modular induction linac will allow us to vary dose rates and pulse lengths over a large parameter range (e. g. with pulse lengths of sub-ns to hundreds of ns and dose rates of a few mA/cm$^2$ to >1 A/cm$^2$).

Thermionic alumino-silicate emitters are known to generate very pure, high brightness, single charge state alkali ion beams [6]. In the present study, we did not measure the elemental composition of the ion beam pulses. Pulses from the injector show indications of impurities present (predominantly K$^+$), mostly during early stages of ion source activation. The accelerator configuration with alternating pulsed voltage gaps and solenoids is very restrictive to the ion mass and charge state and quickly filters out any impurities through time of flight and magnetic solenoid transport optics. From measurements of the ion pulse phase space and from simulations we estimate that the divergence angle of the ions when they impinge on targets is about 0.3 to 1°. Use of longitudinal drift compression of ion pulses with velocity tilts results in a spread of ion kinetic energies in single pulses. Typical values of the head-to-tail energy variation are 20-30% of the average beam energy during compression. The kinetic energy range of ions in the compressed pulse in Figure 2 was about 230 to 320 keV.

In order to use the intense, short ion beam pulses of NDCX-II to access defect dynamics, the beam pulse has to be followed by a suitable probe that can track characteristic signatures of, for example, point defects or clusters of interstitials in irradiated samples. In Table 2, we list a series of diagnostics approaches. In the example of x-ray emission spectroscopies, the ion beam pulse acts as the pump, and a short x-ray pulse is the probe pulse that tracks, e. g., the density and chemical coordination of interstitials for a series of time periods following the ion pulse. This probe could ionize inner shell electrons of selected elements after selected delay periods and the resulting x-ray emission could be measured time-resolved with energy dispersive detectors. Chemical shifts would map the x-ray energy to the bonding configuration of the probed atoms after a selected delay following the ion beam pulse. Combining short ion beam pulses with short laser pulse probes has been suggested for studies of defect dynamics in nuclear materials at the Swiss x-ray FEL [1]. The pulse repetition rate of an FEL is critical in order to efficiently track multi-scale phenomena, such as defect dynamics, and many other problems in meso-scale science from picosecond to many seconds. A prominent design of a high repetition rate x-ray FEL facility, the Next Generation Light Source (NGLS), features a 1 MHz repetition rate for high brightness x-rays [7]. An NDCX-II type accelerator could be integrated with one of the experimental end stations of an FEL facility like NGLS where a full suite of time resolved x-ray emission spectroscopy and diffraction techniques could be implemented for studies of radiation effects in a broad range of materials.

Other probe diagnostics include time resolved measurements of back scattered or transmitted ions, which would track defect evolution during an ion pulse or during an ion pulse sequence. Implementing time resolved ionoluminescence and electrical resistivity [8] measurements is straightforward, while time resolved and energy dispersive measurements of intense ion pulses is much more challenging [9]. *Ex situ* measurements using standard structural probes such as Transmission Electron Microscopes (TEM), or Rutherford Backscattering with channeling analysis (ch-RBS) can complement in situ probes.

The unique, enabling characteristics of NDCX-II are short pulses with a high density of ions. Short pulses enable access to dynamics, while a high density of ions increases signal intensities compared to the impact of single atomic ions or clusters. The current design goal of NDCX-II is to deliver 20 to 50 nC ($1.2 \times 10^{11}$ to $3 \times 10^{11}$) $Li^+$ ions in ~0.6 ns long pulses. This is sufficiently short to access warm dense matter states, since the hydrodynamics expansion time of a heated foil sample is given roughly by the thickness of the heated foil (e. g. the range of 1.2

MeV Li$^+$ ions in aluminum is about 3 μm) divided by the speed of sound (e. g. 6300 m/s for Al). For defect dynamics studies, pulse lengths shorter than 100 ps can become accessible with refinement of pulsed ion source and drift compression techniques.

One intriguing variation on the "pump-probe" theme is to derive two separated ion pulses form a single extracted pulse and thus implement an ion beam based pump-probe experiment. The first pulse would initiate a series of collision cascades and defects will evolve during a (variable) delay time. The second ion pulse then interrogates a sample lattice where a given number of defects has recombined and some are still present in a series of configurations such as single point defects, clusters of defects or captured at an interface. The second ion pulse could also be followed by a third pulse from an auxiliary probe such as an x-ray beam, as discussed above. An alternative to this auxiliary probe is to derive a diagnostic signature from the second ion pulse where, e. g., the signal of transmitted or backscattered ions could track the concentration of defects present during the second probe pulse. As the second ion pulse also induces defects, this ion beam pump-probe diagnostic will have to be carefully calibrated and compared to results from single ion pulses. Yet, the tuning flexibility of NDCX-II offers the possibility to implement this mode of access to defect dynamics. In Figure 4 we show an example of an (imperfect) ion pulse pump-probe sequence. A single pulse is split into two separate pulses. However, refinement of pulsed waveform tuning is required to achieve better pulse separation and to reduce the ion beam intensity between pulses to zero.

**3. Discussion of defect dynamics studies enabled by short ion beam pulses**

We now briefly outline the problem of radiation induced defects and their dynamics in solids and describe one specific experiment that we conducted with ion beam pulses of different pulse lengths from NDCX-II. In this experiment we aim at demonstrating that we can access characteristic time scales of defect recombination in silicon crystals by varying the duration of pulses used for the implantation of lithium ions from 50 ns to 600 ns.

The dynamics of radiation induced defects in materials becomes experimentally accessible with repeatable, well-defined short pulse ion beams in pump-probe type pulse sequences. When energetic ions impinge on a target material they transfer kinetic energy in collisions to target electrons and nuclei. Primary knock-on atoms are dislodged from lattice sites and transfer their kinetic energy to other target atoms during a collision cascade on a picosecond time scale. Most of the ensuing vacancies and interstitials recombine rapidly while the cascade cools. But some defects remain and can diffuse through the disordered lattice until they also recombine or are bound at defect sinks such as interfaces in structured materials, or point defects can accumulate forming extended defects such as voids and dislocation loops [12]. Defect dynamics is a multi-scale problem, spanning time scales from picoseconds to seconds and years. This multi-scale character makes defect dynamics notoriously difficult to track in models and simulations. Multi-scale modeling has to rely on experimental input. But benchmarking of defect dynamics especially on short time scales has been very challenging [11, 13]. Understanding multi-scale defect dynamics promises to guide materials optimization e.g. for the development of improved structural materials, for advanced materials in high radiation environments (e.g. as part of the nuclear fuel cycle or in future fusion reactors) [14] and also to tailor desired properties through defect engineering [15]. An example of the latter are color centers in diamond, specifically the nitrogen-vacancy center, which has received much attention due to evolving applications in quantum information processing [16] and high resolution magnetometry [17]. Yet, reliable formation of NV-centers remains very challenging and our understanding of NV-formation dynamics is incomplete [18].

Dose rate effects have been studied for many years [19-21], revealing in some cases drastic differences in damage accumulation when the same ion fluence was delivered at different rates or during varying ion pulse durations. This has enabled quantification of characteristic damage annealing times on the microsecond to millisecond time scale [20, 21]. One standard *ex situ* diagnostic for dose rate effects during ion implantation is the depth profiling of the implanted ions [20, 22]. Trailing ions in a pulse probe a more disordered lattice, which affects their trajectories, e.g. reducing channeling in crystalline target materials. Use of very intense, short ion beam pulses has also been explored for semiconductor processing, such as simultaneous implantation and annealing [23-25]. In NDCX-II, the ion beam is transported through a beam line with a series of acceleration and beam shaping elements, leading to pure $Li^+$ beams with tunable angular divergence. The critical angle for axial channeling of lithium ions at ~170 keV in silicon (100) is ~1.5° [26], and we estimate the beam divergence in our $Li^+$ pulses to be ~0.3 to 1°. Consequently, a fraction of lithium ions will transverse a silicon lattice on channeling trajectories for irradiations under normal incidence. Ions that impinge on random trajectories damage the lattice at higher rates than ions on channeling trajectories. Ions on channeling trajectories that arrive later in the pulse or in a delayed probe pulse probe the damage that is still present from earlier collision events. By changing the pulse lengths or the delay between pump and probe pulses, it then becomes possible to probe defect recombination on the time scale given by the pulse lengths or the selected delay between pump and probe pulses. In the examples of Figures 2 and 4, pulse lengths are a few tens of ns, while the design goal of NDCX-II is pulse compression to 0.6 ns. Here, the beam spot can be controlled to either advance or suppress sample heating. In the case of lithium ions, resulting range profiles can be impacted by fast diffusion at elevated sample temperatures. In silicon crystals this can be expected for energy fluences >0.1 J/cm$^2$ [23-25].

In Figure 5, we show results from implantation of lithium ions into silicon (100) samples at NDCX-II with two different ion pulse lengths. First, we implanted Li$^+$ in 600 ns long pulses with a current density of 5 mA/cm$^2$. Here, no drift compression was applied and the kinetic energy was 135 keV. The Secondary Ion Mass Spectrometry (SIMS) depth profile [27] shows two peaks, from ions traversing the silicon lattice on random and on channeling trajectories. The peak form channeled ions is at a depth ~1.6 times deeper that the peak from ions on random trajectories. We also found about equal intensities of random and channeled ions in SIMS measurements of control samples that had been implanted with a low dose rate (<10 µA/cm$^2$) and with a beam divergence angle of about 0.3 to 1°, well below the critical angle of 1.5° for axial channeling into Si (100). We then implanted silicon crystals with much shorter, i. e. 50 ns, pulses and a corresponding dose rate of 38.5 mA/cm$^2$. Here, acceleration and drift compression was applied resulting in a kinetic energy, $E_{kin}$, of 170 keV. The SIMS depth profile for the short pulse implant shows a strong reduction of the channeling peak. We did not measure the divergence angle of the ions in these NDCX-II shots *in situ* but from simulation results and beam phase space measurements under similar tuning conditions using a pepper pot technique we estimate it to be about 0.3 to 0.5°. Given a divergence angle that is comparable for the long and short ion pulses the absence of a channeling peak for the short pulse implantation shows that the lattice disorder during 50 ns pulses was much higher than during the lower dose rate implantation with 600 ns pulses.  Incident lithium ions create vacancies and interstitials, which cascade, diffuse and recombine on characteristic time scales.  The absence of the channeling peak for the implants with high peak currents suggests that most of the defects, such as interstitials, that were formed by ions in the leading part of the pulse were still present when trailing ions transverse areas close to areas impacted earlier in the pulse.  But during the 600 ns pulses, defects have enough time to recombine before impacted sample areas are hit by trailing ions and ions see on average a much less defective lattice, resulting in a much larger fraction of channeled ions.  The drastic reduction of channeling when decreasing the pulse lengths from 600 ns to 50 ns thus indicates that a characteristic defect recombination time in silicon at room temperature is of order 100 ns for irradiation with 170 keV $Li^+$.  Recently, ion beam pulsing on a millisecond time scale was used to elucidate a characteristic time scale for damage recovery of radiation damage in silicon at room temperature.  Myers et al. recently reported a dynamic annealing time of ~6 ms [21] for pulsed implantations of silicon crystals with 500 keV $Ar^+$ (7° tilt) with dose rates up to a few $\mu A/cm^2$ (~$1.5x10^{13}$ ions/cm$^2$/s).  This time is much longer than the 100 ns time scale that we extracted here.  Our results are preliminary in the sense that we did not measure the angular spread of the pulsed lithium ion beam *in situ*.  The discrepancy to Myers et al. might stem from the over ten times higher defect density induced by 500 keV $Ar^+$ ions vs. the 170 keV $Li^+$ used here.  From SRIM [28] we can estimate a damage rate of ~5200 displacements per argon ion over a depth of 600 ns, compared to ~550 displacements over ~1 $\mu m$ for the lower energy lithium ions.  Also, the dose rates in our implantations were over 1,000x higher than those used by Myers et al.  We are addressing this discrepancy in follow up studies with a series of pulse lengths, beam energies and dose rates.  In fact, a detailed understanding of defect density effects on dynamic annealing rates and efficiencies is critical for the application of ion beams in accelerated testing of advanced nuclear materials [29]. Range profiles and in particular the ratio of channeled to non-channeled ions implanted into crystal targets are highly sensitive to the presence of lattice disorder during the time of ion implantation and this can be probed through standard SIMS depth profiling. We note that SIMS with a sensitivity of about $5\times10^{13}$ Lithium atoms/cm$^{-3}$ allows tracking of depth profiles from single lithium ion shots with only ~0.6 nC/cm$^2$ ($4\times10^9$ ions/cm$^2$). Beyond this *ex situ* and somewhat indirect diagnostics, we envision implementation of *in situ* diagnostics techniques for sensitive multi-scale tracking of defect dynamics as discussed above.

## 4.       Conclusions and Outlook

The development of short, intense ion beam pulses with high degrees of pulse shape control at the Neutralized Drift Compression eXperiment (NDCX-II) provides a novel tool for studies of defect dynamics in solids. Results from beam shaping experiments have validated the accelerator design concept and we have achieved pulse compression to 20 ns at 320 keV with $8\times10^{10}$ ions/pulse (FWHM). Results from initial pulse shaping experiments demonstrate a basic capability for tuning of pump-probe type ion pulse sequences. Analysis of depth profiles from implantation of lithium ions into silicon samples shows strongly reduced channeling during ion implantation with 50 ns pulses vs. 600 ns pulses, indicating a defect recombination time of about 100 ns under these irradiation conditions. Ion channeling is very sensitive to lattice damage and ex situ analysis of depth profiles using SIMS is one readily accessible technique for probing of defect dynamics in crystalline solids. We discuss the coupling of short ion beam "pump" pulses with auxiliary probe pulses and diagnostics for future pump-probe experiments. Here, coupling of short pulse ion beams with an x-ray FEL promises to provide great versatility for studies of multi-scale dynamics of radiation effects in a broad range of materials (solids, soft matter, and liquids).

**Acknowledgements**

This work was supported by the Office of Science of the US Department of Energy and by the Laboratory Directed Research and Development Program at Berkeley Lab under contract no. DE-AC02–05CH11231. AM was supported by the Center for Defect Physics, an Energy Frontier Research Center funded by the U.S. Department of Energy, Office of Science, Basic Energy Sciences.

Table 1: Parameters for lithium ion pulses from NDCX-II.

| Ion Pulse Parameter | status (March 2013) | Design goals [WW] |
|---|---|---|
| Pulse length (FWHM) | 20 to 600 ns | 0.6 ns |
| kinetic energy | 0.13 to 0.35 MeV | 1.2 MeV |
| ions per pulse | $\sim 10^{10}$ to $5 \times 10^{10}$ | $3 \times 10^{11}$ |
| Ion beam spot size (FWHM) | 1 to 5 $cm^2$ | $\sim 0.6$ $mm^2$ |
| Beam energy fluence | $\sim 1$ $mJ/cm^2$ | 5 to 10 $J/cm^2$ |

Table 2: List of in situ and ex situ diagnostic techniques that we envision can be used to track defect dynamics in materials exposed to short, intense ion beam pulses in single pulses or for sequences of pump-probe pulses with variable delay.

| Diagnostics approach | Instrumentation | Time scales probed | Samples | |
|---|---|---|---|---|
| X-ray spectroscopic methods (e. g. emission spectroscopy) | High brightness x-ray pulses from an FEL [1, 7] | ps to seconds | solids, soft matter, liquids | *In situ* |
| Ionoluminescence | Optical spectrometry with fast streak camera | ns to ms | semiconductors and insulators | *In situ* |
| Energy, angular distributions of transmitted or backscattered ions | Mass spectrometry [9] | few ns to several hundred ns | Transmission: thin crystals; Backscattering: bulk crystals | *In situ* |
| Electrical measurements | Electrometry [8] | ns to seconds | Patterned metal and semiconductor thin films with electrodes | *In situ* |
| Depth profiling of implanted ions | Secondary Ion Mass Spectrometry (SIMS) | few ns to several hundred ns | single crystals (for channeling) | *Ex situ* |
| Structural analysis | Transmission Electron Microscopy (TEM) [10, 11] | ns/ms to days after irradiation | single crystal and poly-crystalline samples | *In situ/Ex situ* |
| Structural analysis | Channeling RBS [21] | days after irradiation | single crystals | *Ex situ* |

**Figure Captions**

Figure 1: Schematic of the NDCX-II accelerator. The total length from injector (left) to target chamber (right) is 12 m.

Figure 2: Lithium ion beam current traces for a single, compressed pulse. The pulse width is 20 ns (FWHM) with 13 nC ($8 \times 10^{10}$ ions), $E_{kin}$=320 keV. The insert shows a 430 ns long pulse (FWHM) with 25 nC ($1.5 \times 10^{11}$ ions), $E_{kin}$=135 keV, from the injector with no active induction cells.

Figure 3: Gated CCD-camera image of ion beam induced luminescence from a single, compressed pulse impinging on a thin alumina scintillator with a beam diameter of 12 mm (FWHM).

Figure 4: Ion beam current traces where a single pulse was shaped into a double peak "pump-probe" type structure. Both the pump and the probe pulses had a length of ~70 ns (FWHM) and the pump-probe delay was 56 ns.

Figure 5: SIMS depth profiles of $^{7}Li^{+}$ ions implanted into silicon (100) under normal incidence. Black: pulse length: 600 ns, $E_{kin}$=135 keV, dose rate: 5 mA/cm$^2$, ten shots, fluence: $2 \times 10^{11}$ cm$^{-2}$, Red: pulse length 50 ns, $E_{kin}$=170 keV, dose rate: 38.5 mA/cm$^2$, ten shots, fluence: $1.2 \times 10^{11}$ cm$^{-2}$.

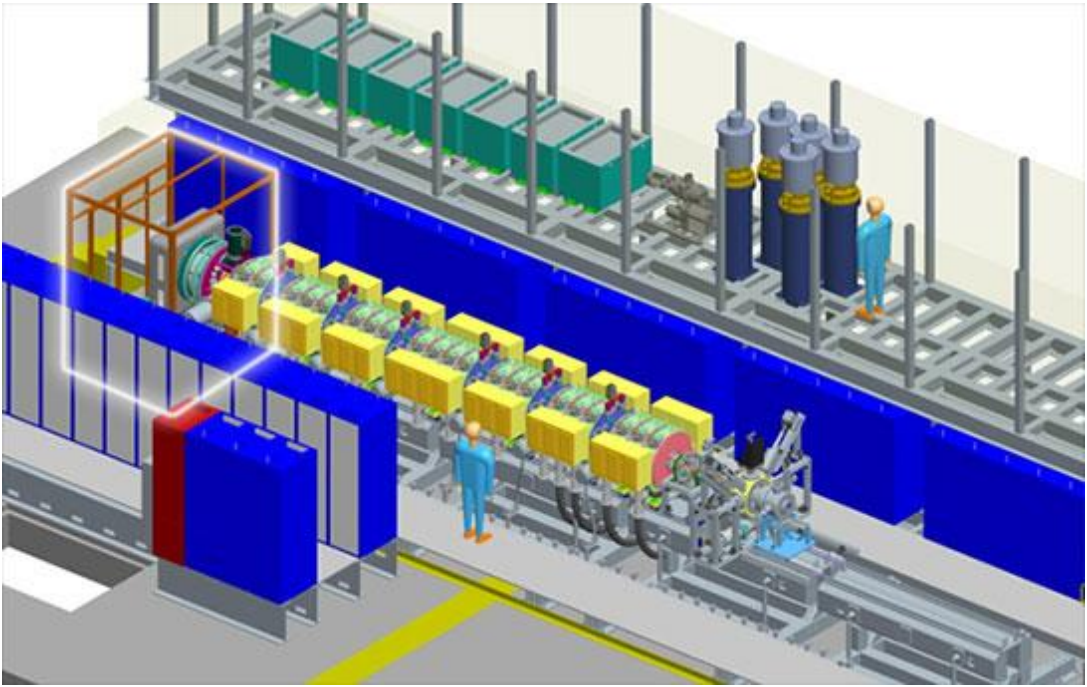

Figure 1

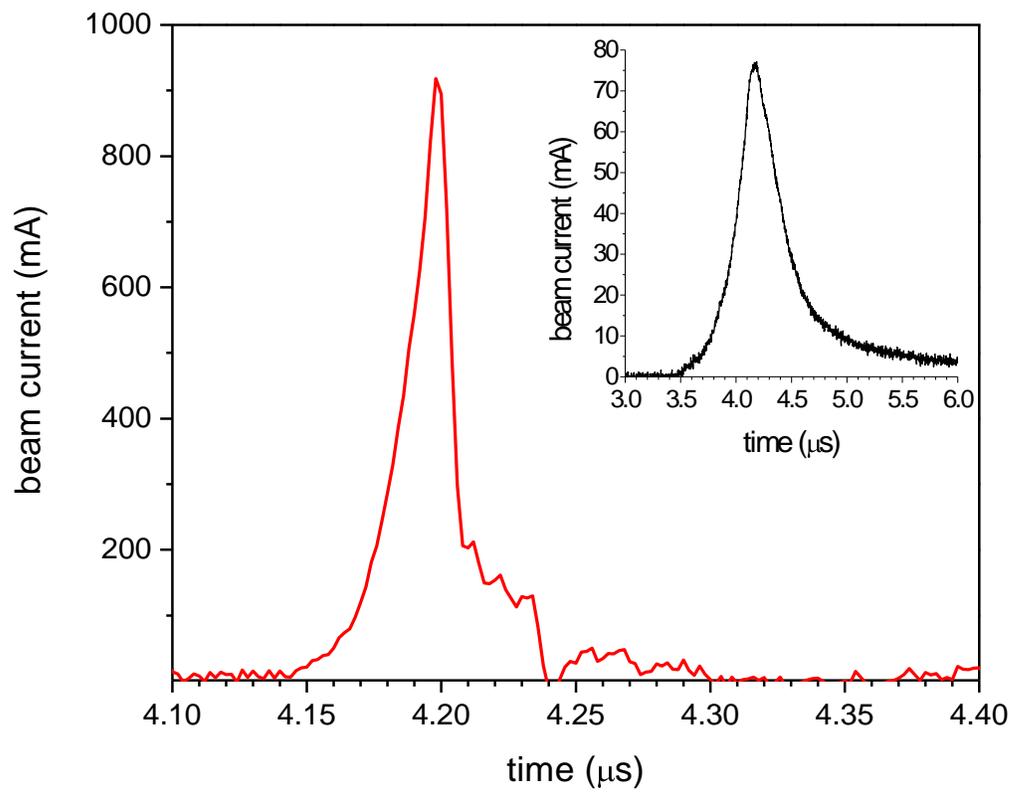

Figure 2

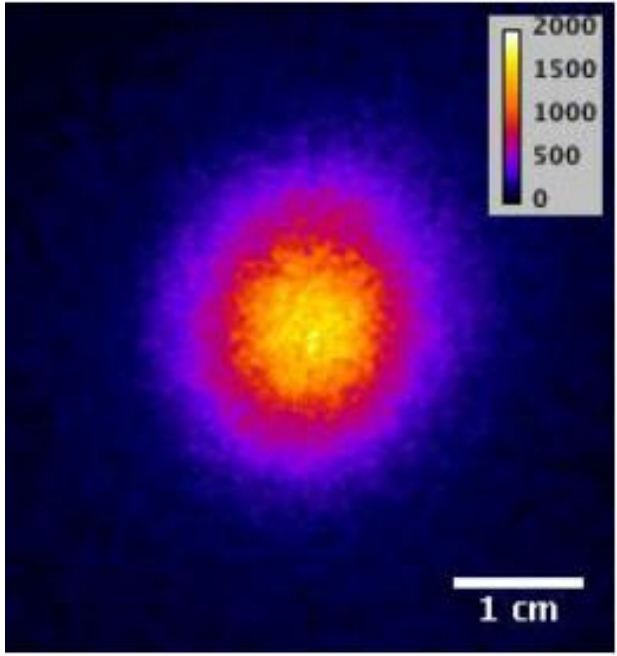

Figure 3

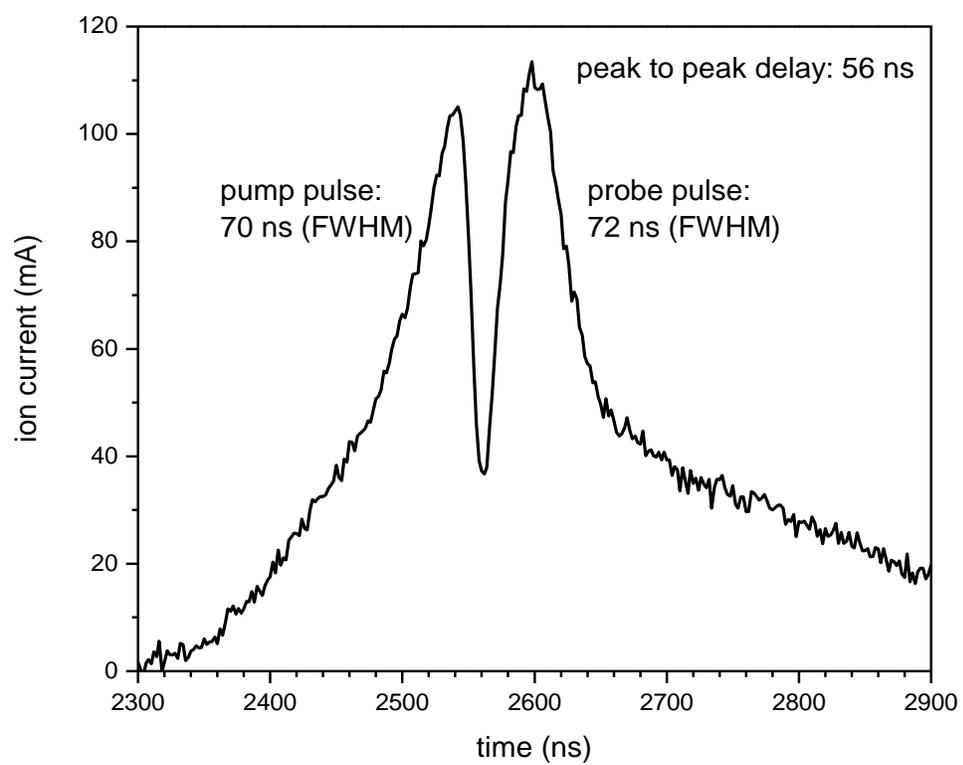

Figure 4

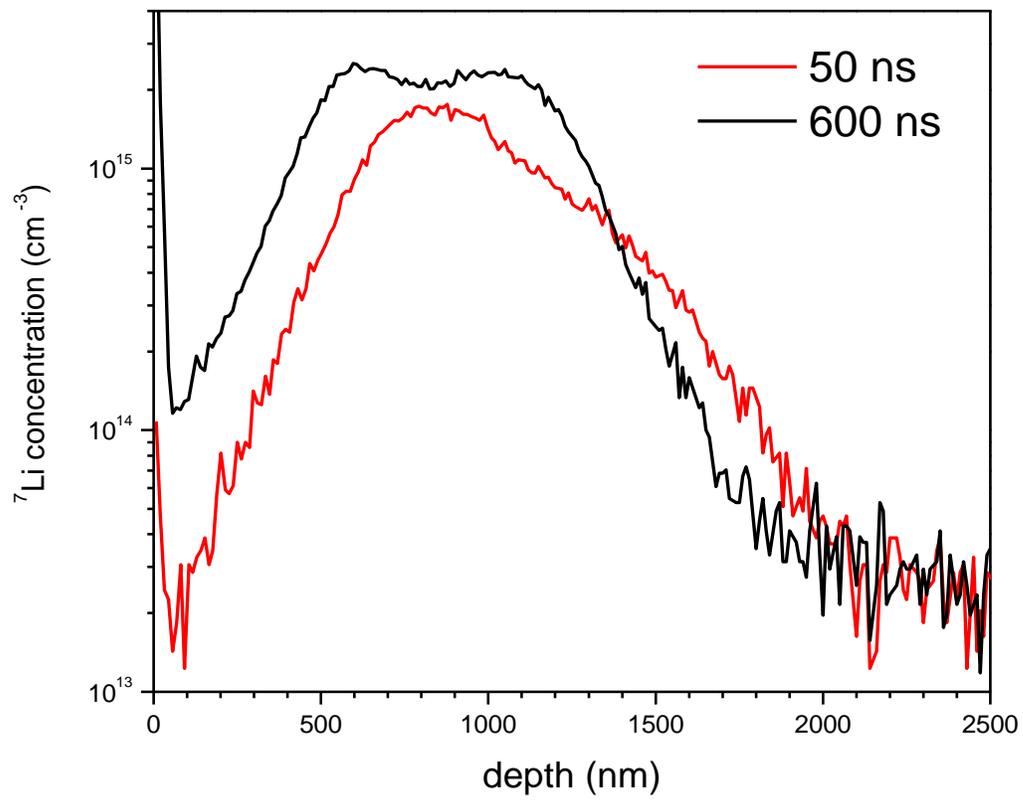

Figure 5